\documentclass[twocolumn]{aastex63}

\graphicspath{{./}{figures/}}

\received{June 12, 2020}
\revised{January 22, 2021}
\accepted{January 22, 2021}

\shorttitle{Black Hole and Galaxy Coevolution at $z\sim1.4$ in SXDF}
\shortauthors{Setoguchi et al.}

\begin{document}

\title{Black Hole and Galaxy Coevolution in Moderately Luminous 
Active Galactic Nuclei at $z\sim1.4$ in SXDF}

\correspondingauthor{Kenta Setoguchi}
\email{setoguchi@kusastro.kyoto-u.ac.jp}

\author[0000-0001-5353-7635]{Kenta Setoguchi}
\affil{Department of Astronomy, Kyoto University, Kitashirakawa-Oiwake-cho, Sakyo-ku, Kyoto 606-8502, Japan}

\author[0000-0001-7821-6715]{Yoshihiro Ueda}
\affiliation{Department of Astronomy, Kyoto University, Kitashirakawa-Oiwake-cho, Sakyo-ku, Kyoto 606-8502, Japan}

\author[0000-0002-3531-7863]{Yoshiki Toba}
\affiliation{Department of Astronomy, Kyoto University, Kitashirakawa-Oiwake-cho, Sakyo-ku, Kyoto 606-8502, Japan}
\affiliation{Academia Sinica Institute of Astronomy and Astrophysics, 11F of Astronomy-Mathematics Building, AS/NTU, No.1, Section 4, Roosevelt Road, Taipei 10617, Taiwan}
\affiliation{Research Center for Space and Cosmic Evolution, Ehime University, 2-5 Bunkyo-cho, Matsuyama, Ehime 790-8577, Japan}

\author[0000-0002-2651-1701]{Masayuki Akiyama}
\affiliation{Astronomical Institute, Tohoku University, 6-3 Aramaki, Aoba-ku, Sendai, Miyagi 980-8578, Japan}



\begin{abstract}

We investigate the relation of black hole mass versus host stellar mass
and that of mass accretion rate versus star formation rate (SFR) in
moderately luminous ($\log L_{\rm bol} \sim 44.5-46.5\ {\rm erg\ s^{-1}}$), X-ray selected
broad-line active galactic nuclei (AGNs) at $z=1.18-1.68$ in the
Subaru/XMM-Newton Deep Field. The far-infrared to
far-ultraviolet spectral energy distributions of 85 AGNs are reproduced
with the latest version of Code Investigating GALaxy Emission ({\tt
CIGALE}), where the AGN clumpy torus model {\tt SKIRTOR} is
implemented. Most of their hosts are confirmed to be main sequence star-forming galaxies. We find that the mean ratio of the black hole mass
($M_{\rm BH}$) to the total stellar mass ($M_{\rm stellar}$) is $\log M_{\rm
BH}/M_{\rm stellar} = -2.2$, which is similar to the
local black hole-to-bulge mass ratio.
This suggests that if the host galaxies of these moderately luminous
AGNs at $z\sim1.4$ are dominated by bulges, they already established
the local black hole mass-bulge mass relation; if they are disk
dominant, their black holes are overmassive relative to the bulges. AGN bolometric luminosities and SFR
show a good
correlation with ratios higher
than that expected from the local black hole-to-bulge mass relation, 
suggesting that these AGNs are in a SMBH-growth dominant phase.

\end{abstract}


\section{Introduction} \label{sec:intro}

The evolution of supermassive black holes (SMBHs) and their host
galaxies is one outstanding question in astrophysics. In the
local universe ($z<1$), a tight correlation between SMBH mass
($M_{\rm{BH}}$) and galactic classical bulge mass ($M_{\rm{bulge}}$) has
been discovered \citep[e.g.,][]{Magorrian,Marconi03,Haring,Gultekin,Kormendy}. This correlation
indicates coevolution between SMBHs and their host galaxies \citep[e.g.,][]{Kormendy}. 
A key population to unveil the origin of the coevolution is active galactic nuclei (AGNs) at cosmic noon ($z\sim1-3$), when the bulk of the growth of SMBHs
and galaxies took place \citep[e.g.,][]{Madau,Ueda14,Aird15}.
Using multiwavelength data in deep survey fields, 
many authors investigated the relation between 
the AGN luminosity (or mass accretion rate onto the SMBH) and the star
formation rate (SFR) of the host galaxy, which 
represent the mass growth rates of the SMBH and galaxy, respectively
\citep[e.g.,][]{Rosario12,Stanley15,Yang17,Ueda18,Yang19,Aird19,Stemo20}.
The relation between the SMBH mass ($M_{\rm{BH}}$) and host stellar mass ($M_{\rm{stellar}}$)
has been also
studied for broad-line AGNs whose black hole masses were 
determined using the broad-line widths and continuum luminosities (e.g.,
\citealt{Jahnke09,Merloni10,Suh20}). 
However, due to observational difficulties, 
studies based on spectroscopically measured $M_{\rm{BH}}$ are still limited.
It is thus important to systematically study the relations among
$M_{\rm{BH}}$, AGN luminosity (or its ratio to $M_{\rm{BH}}$, the
Eddington ratio $\lambda_{\rm Edd}$), $M_{\rm{stellar}}$, and SFR using
a highly complete AGN sample at cosmic noon.

The Subaru / \textit{XMM-Newton} Deep Field (SXDF; \citealt{Sekiguchi})
is one of the best-studied deep multiwavelength survey fields.
\cite{Ueda08} presented the X-ray source catalog from the original 7
XMM-Newton pointings covering an area of 1.14 deg$^2$\footnote{See
\cite{Kocevski} for the Chandra catalog, which covers a 0.33 deg$^2$
area with deeper flux limits.}, whose multiwavelength (radio, mid-IR to
far-UV) properties were studied by \cite{Akiyama}. The spectroscopic or
photometric redshifts were available for all the objects. \cite{Nobuta}
estimated the AGN luminosity at the rest-frame 3000 \AA ($L_{\lambda
3000}$) from the optical spectra, optical photometries, or X-ray
luminosities and derived $M_{\rm{BH}}$ and Eddington ratios of
broad-line AGNs at $z\sim1.4$ from the line widths of Mg II or H
$\alpha$ lines and continuum luminosities ($L_{\lambda 3000}$) in the
spectroscopic survey data\footnote{See also \cite{Oh} for additional
measurements of $M_{\rm{BH}}$ at different redshifts.}. The rich
multiwavelength photometric datasets and the highly completep catalog of
black hole masses available for broad-line AGNs at $z=1.18-1.68$ provide
us with an ideal opportunity to study the relations among physical
properties of the AGN and host galaxies at these redshifts.

The structure of this paper is as follows. We describe the details of
sample selection and the method of the spectral energy distribution
(SED) fitting in Section \ref{sec:data}. To better estimate the SFRs, we
add far-infrared (FIR) photometric data by cross-matching with the \textit{Herschel} Multi-tiered Extragalactic
Survey (HerMES) catalog
\citep{Oliver12}. In Section \ref{sec:results} we perform correlation
analysis between $M_{\rm{BH}}$ and $M_{\rm{stellar}}$ and that between
$\rm{SFR}$ and $L_{\rm{bol}}$ (or $\lambda_{\rm{Edd}}$). The results
are discussed in comparison with previous works. The conclusions are
summarized in Section \ref{sec:conclusion}. Throughout this paper, the
adopted cosmology is a flat universe with $H_0$ = 70 km s$^{-1}$
Mpc$^{-1}$, $\Omega_M$ = 0.3, and $\Omega_{\Lambda}$ = 0.7 .

\section{Data and Analysis} \label{sec:data}

\subsection{Sample Selection} \label{subsec:selection}

To investigate statistical properties of AGNs at $z\sim1.4$, we selected
those at $z=1.18-1.68$ in the SXDF detected with XMM-Newton \citep{Ueda08}
and identified with multiwavelength catalogs \citep{Akiyama}. Among the
117 at $z=1.18-1.68$ broad-line AGNs in the \cite{Akiyama} catalog, black hole
masses of 116 AGNs were measured by \cite{Nobuta}, 
which constitute our parent sample. 
The 3000 \AA \ monochromatic luminosities 
range from $\log
\lambda_{\rm 3000} L_{\lambda 3000} = 43.6$ to $47.3\ \mathrm{erg\ s^{-1}}$ with a median of $44.7\ \mathrm{erg\ s^{-1}}$. 
The Eddington ratios were calculated as $\lambda_{\rm Edd} = L_{\rm
bol}$/$L_{\rm Edd}$, where $L_{\rm bol} = 5.8 \lambda_{3000} L_{\lambda
3000}$ \citep{Richards06} and $L_{\rm
Edd} = 1.25\times10^{38} M_{\rm BH}/M_\odot$. The black hole masses and
Eddington ratios range $\log M_{\rm BH}/M_\odot = 7.2 \sim9.8 $
(median 8.4) and $\log \lambda_{\rm Edd} = -2.06 \sim0.13$ (median $-1.1$),
respectively \citep{Nobuta}.

\subsection{Cross-match with FIR Data from HerMES}
\label{subsec:far-ir}

Since FIR data are important to estimate the SFRs of the host
galaxies, we added FIR photometries obtained by HerMES (\citealt{Oliver12}) to the multiwavelength SEDs in \cite{Akiyama}. Using
the HerMES DR4 catalog in the XMM-LSS field, we performed
nearest-neighbor matching within a search radius of 20\arcsec \ around
the optical counterparts of the X-ray sources. We identified 80, 78, and
65 \textit{Herschel} counterparts at 250, 350, and 500$\micron$,
respectively. For non-detected sources, we assign $3\sigma$ upper limits
of 15.48, 12.72, and 18.48 mJy at 250, 350, and 500 $\micron$,
respectively \citep{Oliver12}.

Since Herschel/SPIRE has a large beam size (full widths at half maximum
(FWHM) of 18.2\arcsec, 24.9\arcsec, 36.3\arcsec\ at 250, 350, and
500\micron, respectively; \citealt{Oliver12}), the measured FIR
photometries may be contaminated by nearby sources. To evaluate
the effect, we checked the image at 24\micron, the closest band to the
Herschel ones, utilizing the Spitzer UKIDSS Ultra Deep Survey (SpUDS)
catalog and the Spitzer Wide-Area
Infrared Extragalactic (SWIRE) legacy survey catalog \citep{Lonsdale03, Lonsdale04}. We searched for 24\micron\ sources within a radius of
18.2\arcsec\ (i.e., the half of the FWHM at 500\micron) around the
position of the optical counterpart.\ We found that 26 out of the 116
objects have multiple counterparts in the 24\micron\ band. For these
sources, we calculated the $3\sigma$ upper boundary of the FIR
photometries and used them as the upper limits.

\ifnum0=1
Using Spitzer UKIDSS Ultra Deep Survey (SpUDS) catalog, we carried out a
nearest-neighbor matching within a search radius of 18.2\arcsec to check
contaminations of our sample at 24\micron, which lies closest to FIR
bands. We find that, among 85 objects of which we estimated SFR and
$M_{\rm stellar}$, 14 objects are contaminated at 24\micron and would be
contaminated similarly in FIR. To minimize the effect of contamination in FIR, we put $3\sigma$ upper limits at 250, 350, and 500\micron and reran a SED analysis for these objects.

A correlation coefficient is $r=-0.061\pm0.201$, $r=0.724\pm0.081$ and $r=0.631\pm0.133$ in $M_{\rm stellar}$ versus $M_{\rm BH}$, SFR versus $L_{\rm bol}$, and SFR versus $\lambda_{\rm Edd}$ ,respectively.
These values are similar to those as discussed in Section 3.3 and 3.4.
Therefore, contaminations in FIR does not have a significant effect on our discussion.
\fi

\begin{figure*}[ht!]
  \plottwo{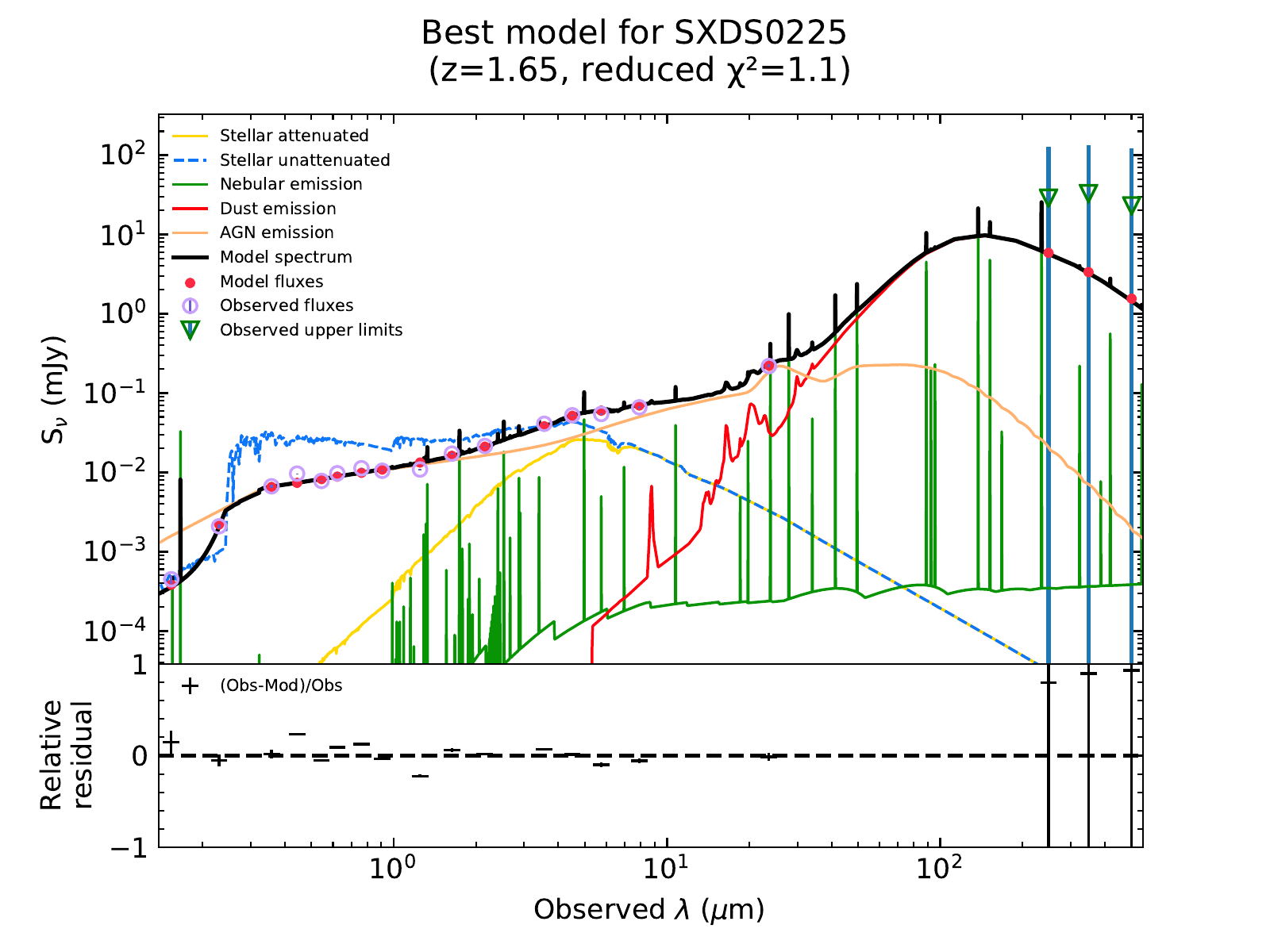}{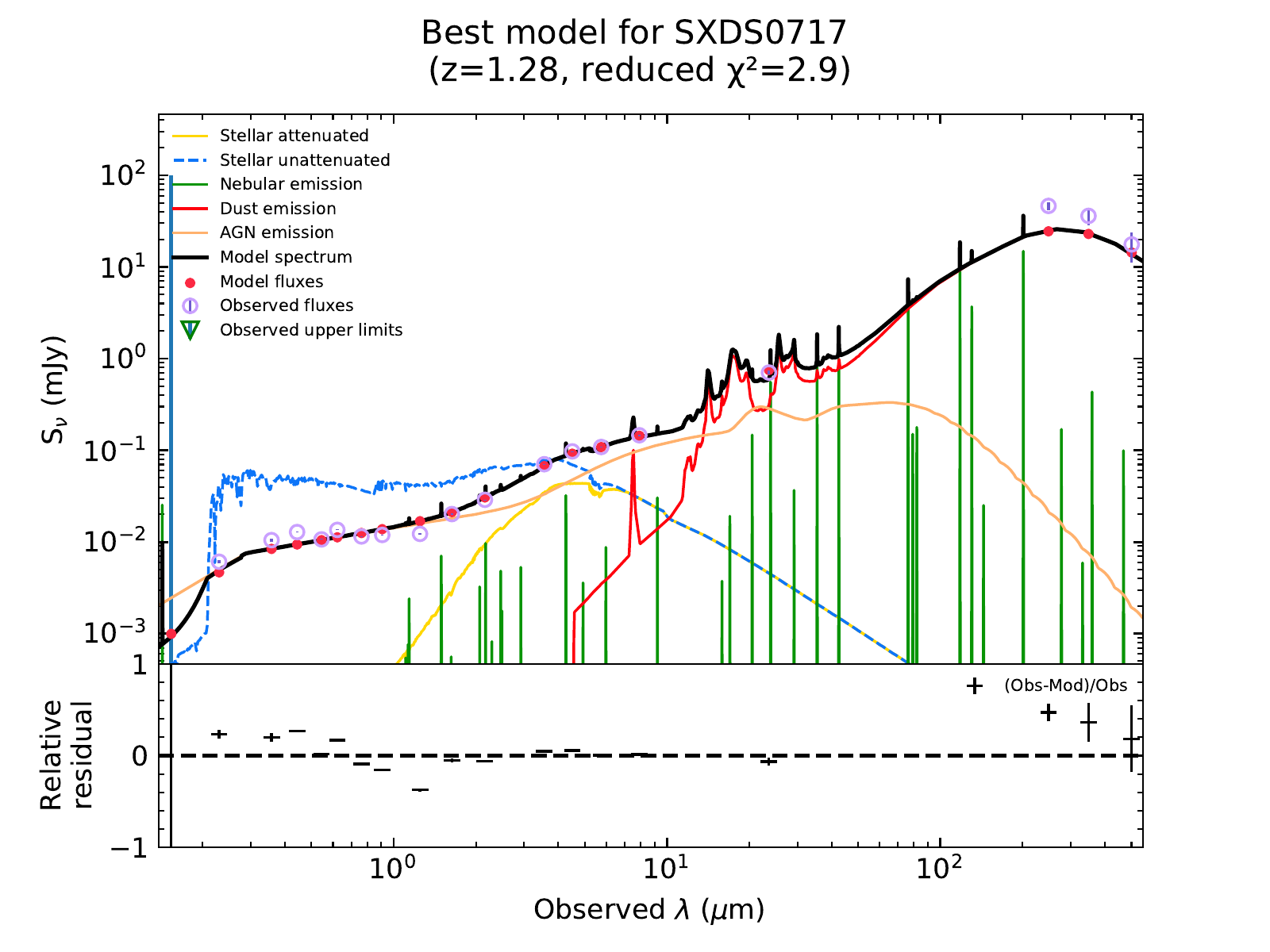}
  \caption{Examples of the SED fitting for our objects with
 CIGALE. SXDS0225 (left) and SXDS0717 (right) are a quasar and a
 Seyfert, respectively. The black solid lines display the best-fit SEDs.\\
  }
  \label{fig:cigale}
\end{figure*}

\subsection{SED Fitting with {\tt CIGALE}}\label{CIGALE}

We performed a multi-component SED fitting to 19 photometries (or their 
upper limits) in the far-IR (Herschel/SPIRE and PACS), mid-IR 
(Spitzer/IRAC and MIPS), near-IR (UKIDSS), optical (Subaru), and 
ultraviolet (GALEX) bands for each object (see Table~1 of \citealt{Akiyama} 
for details except for the far-IR data).

We employed a new version of Code Investigating GALaxy Emission \citep[{\tt CIGALE};][]{Burgarella,Noll,Boquien19}, named {\tt X-CIGALE} \citep{Yang20}, where a 
clumpy two-phase torus model, {\tt SKIRTOR} \citep{Stalevski16}, has been implemented 
as an AGN template.
An advantage of {\tt CIGALE} is that re-emission from dust in the mid-IR 
and far-IR bands is self-consistently calculated by considering the 
energy balance.
Following \citet{Toba19b}, we adopted a star formation history (SFH) of two 
exponentially-decreasing SFRs with different e-folding times: the main stellar population 
($\tau_{\rm{main}}$) and the late starburst one ($\tau_{\rm{burst}}$).
We chose the single stellar population (SSP) model \citep{Bruzual}, 
assuming a Chabrier initial mass function \citep[IMF;][]
{Chabrier}.
The nebular emission model is based on \cite{Inoue}, for which we used 
the default template and parameters.
Dust attenuation was taken into account with the law of \cite{Calzetti}, 
parameterized by the color excess $(E(B-V)_*)$.
The reprocessed IR emission of dust absorbed from UV/optical stellar 
emission is modeled by using dust
templates provided by \cite{Dale}.
For AGN emission, we utilized the {\tt SKIRTOR} model that has 7 
parameters: torus optical depth at 9.7 $\micron$ ($\tau_{\rm 9.7}$), 
torus density radial parameter ($p$), torus density angular parameter 
($q$), angle between the equatorial plane and edge of the torus 
($\Delta$), ratio of the maximum to minimum radii of the torus ($R_{\rm 
max}/R_{\rm min}$), the viewing angle ($\theta$), and the AGN fraction 
in total IR luminosity ($f_{\rm AGN}$).
In order to avoid a degeneracy of AGN templates in the same manner as 
\cite{Yang20}, we fixed $R_{\rm max}/R_{\rm min}$, $\Delta$, and $\theta$ 
that are optimized for type 1 AGNs.
Table \ref{T1} summarizes the free parameters in the SED model.
We note that 
{\tt CIGALE} can handle the upper limits of photometries to perform
Bayesian estimation, utilizing the method by \cite{Sawicki12} (see Section 4.3 in \citealt{Boquien19}, for more detail).

\ifnum0=1
When we determined $M_{\rm stellar}$ and SFR of our objects which could
not be identified counterparts in Herschel, we used upper limits of
bands to constrain the far-infrared re-emission of dust and performed a
Bayesian-like estimate. {\tt CIGALE} utilizes the method based on \cite{Sawicki12} to handle flux densities that can be put only upper limits.
\fi

\begin{table}
\renewcommand{\thetable}{\arabic{table}}
\centering
\caption{Free Parameters used for the SED fitting of our objects with 
{\tt CIGALE}.}
\label{T1}
\begin{tabular}{lc}
\tablewidth{0pt}
\hline
\hline
Parameter & Value\\
\hline
\multicolumn2c{Double exp. SFH}\\
\hline
$\tau_{\rm main}$ [Myr] & 1000, 3000, 4000, 8000 \\
$\tau_{\rm burst}$ [Myr] & 3, 8, 80 \\
$f_{\rm burst}$ & 0.001, 0.1, 0.3 \\
age [Myr] & 1000, 4000, 6000 \\
\hline
\multicolumn2c{Single stellar population \citep{Bruzual}}\\
\hline
IMF & \cite{Chabrier} \\
Metallicity & 0.02 \\
\hline
\multicolumn2c{Dust attenuation \citep{Calzetti}}\\
\hline
$E(B-V)_*$ & 0.01, 0.1, 0.2, 0.3, 0.4, \\
& 0.5, 0.6, 0.7, 0.8, 1.0 \\
\hline

\multicolumn2c{Dust emission \citep{Dale}}\\
\hline
IR power-law slope ($\alpha_{\rm dust}$) & 0.0625, 1.0000, 1.5000, 
2.0000, \\
& 2.5000, 3.0000, 4.0000 \\
\hline
\multicolumn2c{AGN emission \citep{Stalevski16}}\\
\hline
$\tau_{\rm 9.7}$ 			& 	3, 7 		\\
$p$							&	0.0, 1.0	\\
$q$							&	0.0, 1.0	\\
$\Delta$					&	40\arcdeg	\\
$R_{\rm max}/R_{\rm min}$ 	& 	30 			\\
$\theta$					&	30\arcdeg	\\
$f_{\rm AGN}$ 				& 	0.1, 0.3, 0.5, 0.7, 0.9\\
\hline
\end{tabular}
\end{table}

We found that {\tt X-CIGALE} adequately reproduced the SEDs from far-IR
(500$\micron$) to far-UV (1500 \AA) in a majority of objects; 
among the 116 objects, 
92 had reduced $\chi^2<10.0$. 
As a sanity check, we compared the AGN luminosity obtained from {\tt X-CIGALE} and $L_{\rm bol}$ estimated
by \cite{Nobuta}. It is found that 7 out of the 92 objects show significantly smaller AGN luminosities in {\tt X-CIGALE}. Detailed inspection 
suggests that these 7 objects are likely weakly-absorbed AGNs, 
whose SEDs cannot be well reproduced with the current AGN template in
{\tt X-CIGALE}. 
We thus excluded them from the sample and used the remaining 85 objects
in the following analysis. 
Examples of our SED fitting for two 
objects are shown in Figure \ref{fig:cigale}.

\begin{figure*}[ht!] 
  \plotone{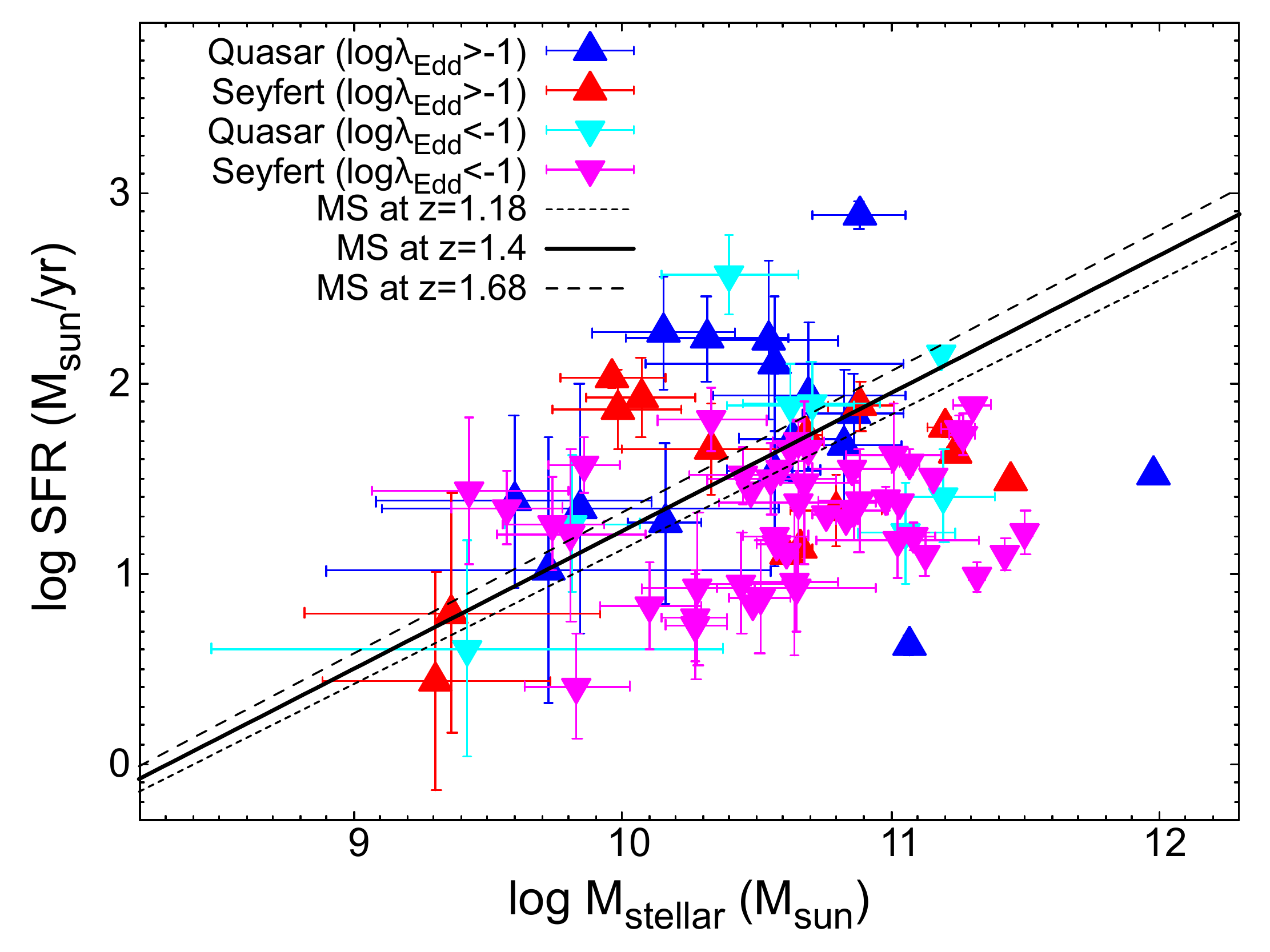}
  \caption{$M_{\rm{stellar}}$ vs. SFR of our objects. The black solid line represents the main sequence relation at $z=1.4$ from \cite{Speagle}. The large blue triangles: quasars ($\log L_{\rm bol}/L_\odot >12$)
 with $\log \lambda_{\rm{Edd}}>-1$. The large red triangles: Seyferts
 ($\log L_{\rm bol}/L_\odot <12$) with $\log
 \lambda_{\rm{Edd}}>-1$. The cyan inverse triangles: quasars with $\log \lambda_{\rm{Edd}}<-1$. The magenta inverse triangles: Seyferts with $\log \lambda_{\rm{Edd}}<-1$.}
  \label{fig:sequence}
\end{figure*}

To check whether their physical properties are reliably estimated given the uncertainties of the photometries, we performed a mock
analysis implemented on {\tt X-CIGALE}. 
It enables us to compare the parameters estimated by {\tt X-CIGALE} using the real catalog with those from the mock
catalog, which {\tt X-CIGALE} produces
from the best-fit SED and
photometric errors for each object (see, e.g., \citealt{Boquien19};
\citealt{Yang20}; \citealt{Toba20c} for more detail). 
We confirmed that our $M_{\rm stellar}$ and SFR values 
adequately agree with those 
obtained from the mock catalog within the errors.

\section{Results and Discussions}\label{sec:results}

\subsection{Results of the SED with {\tt CIGALE}}

In this paper, we focus on the SFR and the total stellar mass
(including those of the bulge [$M_{\rm{bulge}}$] and the galactic disk)
of the host galaxies derived from our SED fitting. We then compare them
with the AGN parameters, $L_{\rm{bol}}$, $M_{\rm{BH}}$, and
$\lambda_{\rm{Edd}}$. For convenience, we divide our sample into four
groups by the bolometric luminosity and Eddington ratio and use different
symbols commonly in all plots. We refer to those with $\log L_{\rm
bol}/L_\odot >12$ and $\log L_{\rm bol}/L_{\odot} <12$ as quasars and
Seyferts, respectively, and those with $\log \lambda_{\rm Edd}>-1$ and
$\log \lambda_{\rm Edd}<-1$ as high Eddington AGNs and low Eddington ones,
respectively.

\subsection{Stellar Mass and SFR}

Figure \ref{fig:sequence} displays the relation between
$M_{\rm{stellar}}$ and SFR of our sample. The stellar masses and SFRs of
our sample range from $\log M_{\rm stellar}/M_\odot$ = 9.3 to 12.0 (median 10.6) and 
from $\log {\rm SFR}/(M_\odot\ {\rm yr}^{-1})$ = 0.4 to 2.9 (median 1.4), respectively. The black solid line
represents the main sequence (MS) relation at $z=1.4$ given by
\cite{Speagle}. As noticed, most of the host galaxies are classified as
star forming galaxies in the MS; only a very small fraction are starburst
galaxies ($>$0.6 dex above the MS line). This result is consistent with
previous studies of X-ray selected AGNs with similar luminosities and
redshifts (e.g., \citealt{Santini12}; \citealt{Yang17}; \citealt{Ueda18}). 

\begin{figure*} 
  \plotone{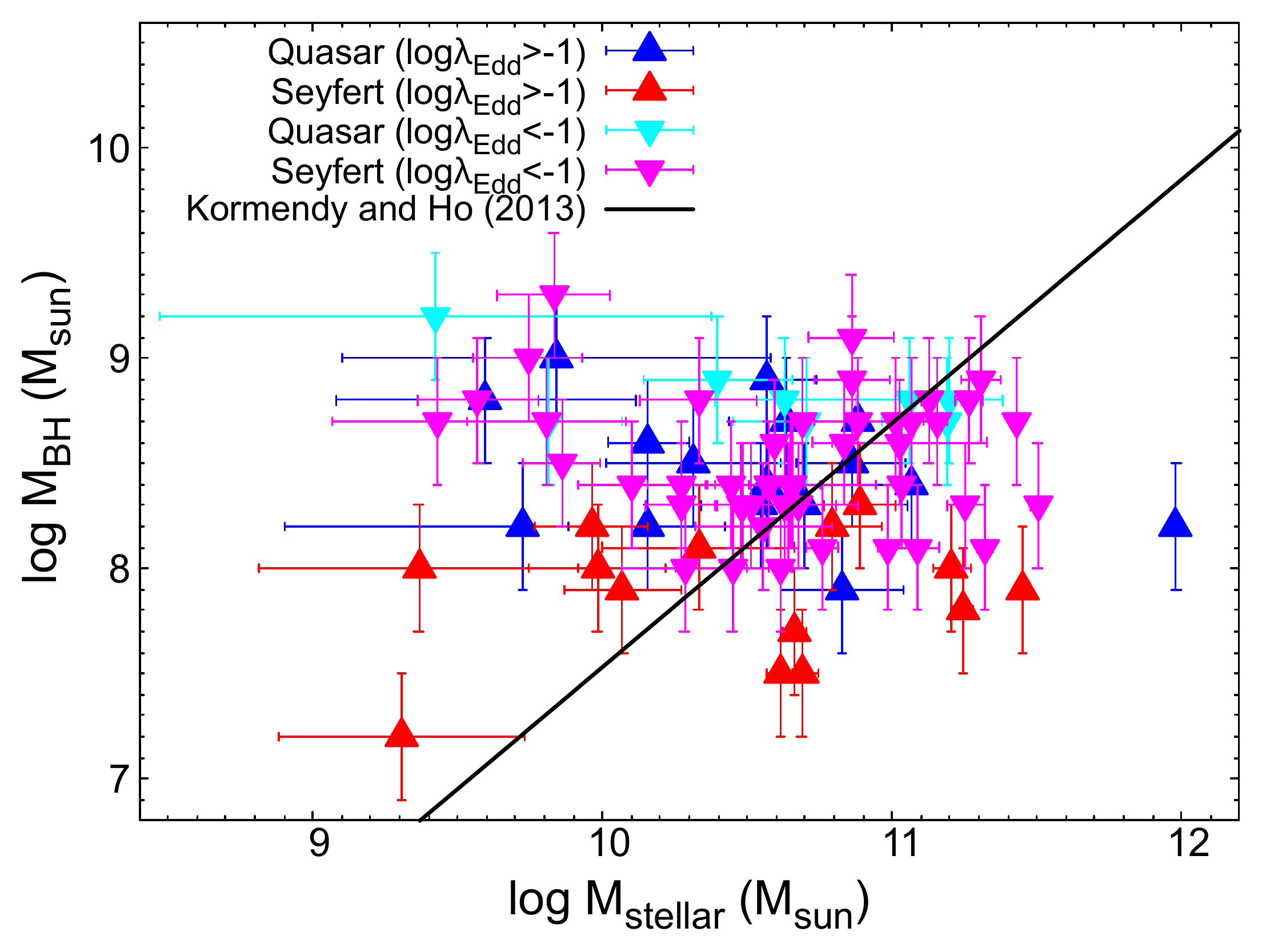}
\caption{
Relation between $M_{\rm{stellar}}$ and
    $M_{\rm{BH}}$. The symbols are the same as in Figure 2. 
The black solid
     line represents the local black hole-to-bulge mass ratio from
     \cite{Kormendy}. 
}  \label{fig:MBHvsMstellar}
\end{figure*}

\subsection{Stellar Mass versus Black Hole Mass}

Figure \ref{fig:MBHvsMstellar} plots the relation between
$M_{\rm{stellar}}$ and $M_{\rm{BH}}$. We perform a correlation analysis
with the method of \cite{Kelly07} where the errors in the two parameters
are taken into account (see also \citealt{Toba_19a} for an application
to a large number of X-ray selected type 1 AGNs). It yields a correlation
coefficient of $r=-0.068\pm0.204$, indicating no correlation. This
would be simply due to the small range of $M_{\rm BH}$ in our sample. A
mean $M_{\rm{BH}}/M_{\rm{stellar}}$ ratio is calculated to be $-2.2$
with a 1$\sigma$ scatter of 0.69.
The black solid line represents the
local $M_{\rm{BH}}$ versus $M_{\rm{bulge}}$ relation for classical
bulges determined by \cite{Kormendy}. On average, our sample has
$M_{\rm{BH}}/M_{\rm{stellar}}$ ratios similar to the local
$M_{\rm{BH}}/M_{\rm{bulge}}$ relation. Since $M_{\rm{stellar}} \geq
M_{\rm{bulge}}$, the $M_{\rm{BH}}-M_{\rm{bulge}}$ relation for disk-dominant galaxies is different from the local one in the sense that
black holes are overmassive.

Our result on the $M_{\rm{BH}}/M_{\rm{stellar}}$ ratio at $1.18<z<1.68$
is consistent with that reported by \cite{Merloni10} 
for type-1 AGNs with similar bolometric luminosities at
$1<z<2.2$ in the COSMOS field (after converting $M_{\rm{stellar}}$ from
a Salpeter IMF to a Chabrier IMF by --0.255 dex). Although \cite{Merloni10}
argued that the $M_{\rm{BH}}/M_{\rm{stellar}}$ ratio evolves with
$(1+z)^{1.15\pm0.13}$ (for a Chabrier IMF) compared with the local 
$M_{\rm{BH}}/M_{\rm{bulge}}$ relation by \cite{Haring}, 
the recent upward correction by \cite{Kormendy} 
in the local $M_{\rm{BH}}/M_{\rm{bulge}}$ ratio by $\sim$0.5 dex would
indicate no evolution from $z=0$ to $z\sim1.4$. More recently, however,
\cite{Suh20} obtained a significantly lower ratio ($\log
M_{\rm{BH}}/M_{\rm{stellar}} \approx -2.7$) than our result
for AGNs at similar redshifts detected in the Chandra COSMOS Legacy
Survey, adopting a Chabrier IMF (i.e., the same as our IMF). Their mean
$M_{\rm{stellar}}$ value is $\sim$0.5 dex larger than our result,
whereas AGN bolometric luminosities of their sample are typically
$\sim$0.5 dex smaller than ours. The reason behind the discrepancy is
unclear; it is most likely caused by different models used in the SED
fitting. Indeed, \cite{Suh19} obtained more massive stellar masses in
type-1 AGNs than in type-2 AGNs at the same redshifts. In this paper,
we do not pursue this issue further, but we always need to bear these
possible systematic uncertainties in mind.

\begin{figure*} 
\gridline{\fig{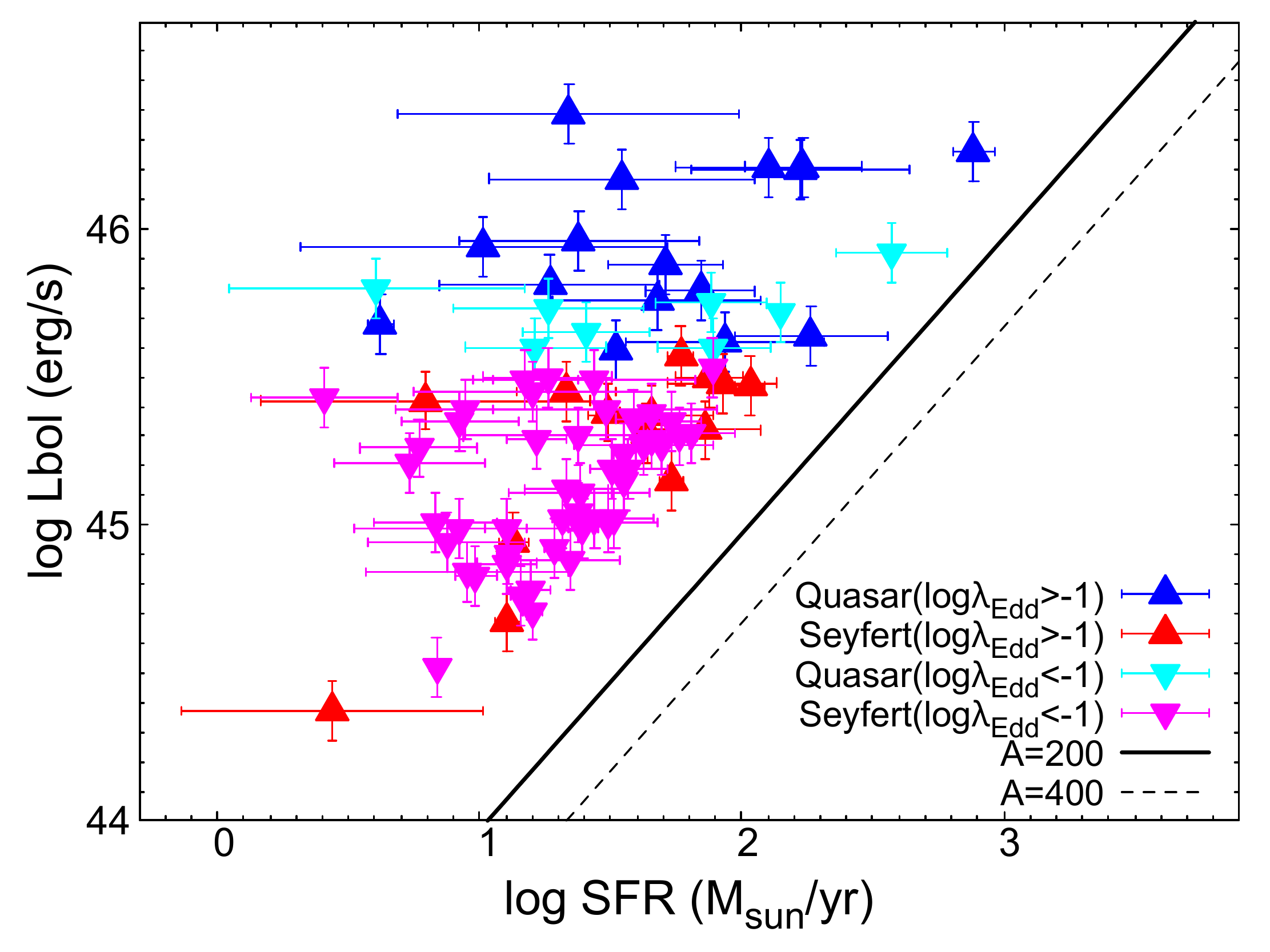}{0.45\textwidth}{(a)}
     \fig{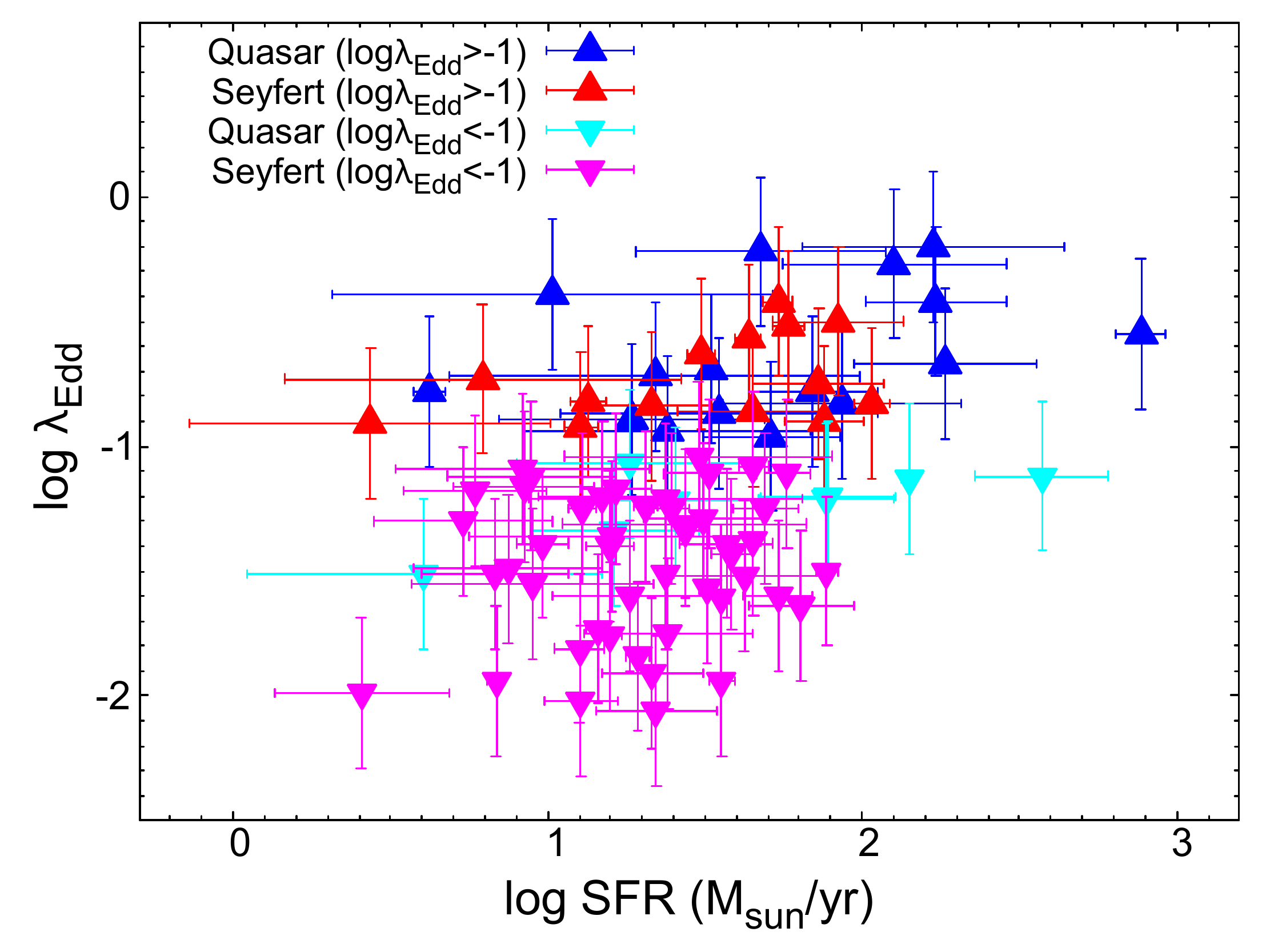}{0.45\textwidth}{(b)} }
\caption{Relations between (a) SFR and $L_{\rm{bol}}$ and (b) SFR and
$\lambda_{\rm{Edd}}$. The symbols are the same as in Figure 2. The
black solid and dashed lines in (a) correspond to equation (1) with
$A=200$ and $A=400$, respectively.}
\label{fig:coevolution}
\end{figure*}

Our result suggests that (1) moderately luminous AGNs at $z\sim1.4$
with bulge-dominant galaxies already established the local
$M_{\rm{BH}}$-$M_{\rm{bulge}}$ relation and that (2) those with disk-dominant galaxies have overmassive SMBHs relative to the bulge stellar
masses. This implies that, in the disk-dominant galaxies, later star
formation in the bulge at $z\lesssim1.4$ caught up the earlier growth of
the SMBH or that the mass of the disk was redistributed into the bulge
by mergers or disk instabilities at $z\lesssim1.4$ (see, e.g., \citealt{Jahnke09}). Such a tendency has been reported by many authors for
luminous AGNs at the same or higher redshifts (see \citealt{Kormendy}
and references therein). It is, however, likely to be subject to selection
bias toward luminous AGNs. It is possible that lower-luminosity AGNs may
have less massive SMBHs \citep[e.g.,][]{Ueda18,Izumi19}.
Hence, the overall picture of SMBH and galaxy
coevolution is still an open question (see Section 3.5 for discussion).

\subsection{SFR versus Black Hole Accretion Rate and Eddington Ratio}

Figure \ref{fig:coevolution} (a) shows the relation between SFR and
$L_{\rm{bol}}$, which represent the time derivatives of
$M_{\rm{stellar}}$ and $M_{\rm{BH}}$ at the observed epoch. We obtain a
correlation coefficient of $r=0.717\pm0.084$, indicating a significant
positive correlation. 
The black solid and dashed lines correspond to the
local $M_{\rm{BH}}$-vs-$M_{\rm{bulge}}$ and
$M_{\rm{BH}}$-vs-$M_{\rm{stellar}}$ relations, respectively, that would
be expected from exactly simultaneous evolution of SMBHs and their host
galaxies. They are given as
\begin{equation}
{\rm SFR} \times (1-R) = A \times L_{\rm{bol}}(1-\eta)/\eta c^2,
\end{equation}
where $R=0.41$ (for a Chabrier IMF) is the return fraction, $A$ the
local star-to-SMBH mass ratio ($A = M_{\rm{bulge}}/M_{\rm{BH}} = 200$ or $A =
M_{\rm{stellar}}/M_{\rm{BH}} = 400$), $\eta=0.05$ the radiation
efficiency, and $c$ the light speed (see \citealt{Ueda18}).
Our objects are located above these lines, indicating
that they are in a SMBH-growth dominant phase.

\citet{Yang17} show that the mean ratio of host galaxy SFR to AGN
luminosity increases with redshift (see also, e.g., \citealt{Stemo20} for
similar results). Generally, a flux-limited sample obtained from a
single survey contains more luminous AGNs at higher redshifts. Then,
even if there were no intrinsic correlation at a given redshift, the
redshift dependence on the SFR-to-$L_{\rm{bol}}$ ratio could drive an
apparent correlation between SFR and $L_{\rm{bol}}$ based on a sample
that covers a wide redshift range. 
Thanks to
the narrow redshift range of our sample ($z=1.18-1.68$), however, the effect is estimated to be negligible compared with the observed scatter in the $L_{\rm{bol}}$-to-SFR ratio. 
\cite{Yang17} and \cite{Stemo20} also report that
the correlation between SFR and mass accretion rate ($L_{\rm{bol}}$)
comes from that between $M_{\rm stellar}$ and $L_{\rm{bol}}$
and the main sequence $M_{\rm stellar}$-SFR relation.
However, we obtain a correlation coefficient between $M_{\rm stellar}$
and $L_{\rm bol}$ of $r=-0.037\pm0.142$ (no significant correlation),
suggesting that the SFR-$L_{\rm bol}$ correlation is a primary one.

We note that this result is subject to selection biases toward luminous
AGNs, similar to the case of the $M_{\rm{stellar}}$ versus $M_{\rm{BH}}$
relation. In fact, deeper Chandra surveys detected a dominant AGN
population whose mass accretion rate-to-SFR ratios are smaller than the
local relation (e.g., \citealt{Yang17}, \citealt{Ueda18}). The 
large scatter
would be explained by time variability of AGN activities
\citep{Hickox14} and/or non co-evolution nature for disk-dominated
systems \citep{Kormendy}. \cite{Yang19} report that the average black
hole accretion rate of all bulge-dominant galaxies at $0.5<z<3.0$ shows
a good correlation with the SFR, whereas disk-dominant ones do not.


We also plot the relation between SFR and $\lambda_{\rm{Edd}}$ in Figure
\ref{fig:coevolution}(b). We obtain a correlation coefficient of
$r=0.623\pm0.136$. This indicates a similarly strong positive
correlation to that between SFR and $L_{\rm bol}$, as expected from a
narrow range of $M_{\rm BH}$ (Figure \ref{fig:coevolution}). To our
knowledge, this is the first report of such correlation based on direct
measurements of $M_{\rm{BH}}$ for $z\sim1.4$ AGNs 
(see \citealt{Zhuang20} for the result for nearby AGNs at $z<0.35$).
\cite{Aird19} show that high SFR galaxies tend to contain more AGNs
with high ''specific'' accretion rates (those divided by their host
stellar masses, instead of the black hole masses). That trend is not
surprising given a moderate scatter between $M_{\rm{BH}}$ and
$M_{\rm{stellar}}$ as shown in Figure \ref{fig:MBHvsMstellar}.

We investigate whether the observed correlation between SFR and
$\lambda_{\rm{Edd}}$ arises from that between SFR and $L_{\rm
bol}$ given the small $M_{\rm BH}$ range. 
Following \cite{Zhuang20}, we
divide our sample by $L_{\rm bol}$ into two groups, 
quasars and Seyferts. 
The correlation coefficients between 
SFR and $\lambda_{\rm Edd}$ are found to be 
$r=0.476\pm0.358$ (quasars) and $r=0.581\pm0.179$ (Seyferts). 
Alternatively, when we divide the sample by $\lambda_{\rm{Edd}}$, we
obtain correlation coefficients between SFR and $L_{\rm bol}$ of
$r=0.634\pm0.167$ and $r=0.743\pm0.115$ for high and low Eddington rate
AGNs, respectively. 
The more significant correlations between SFR and $L_{\rm bol}$ 
than those between SFR and $\lambda_{\rm Edd}$ 
suggest that $L_{\rm bol}$ is likely to be the primary parameter. \cite{Zhuang20} have reached a similar conclusion for
$z<0.35$ AGNs.

\subsection{Evolution of Black Hole-to-Stellar Mass Ratio}

\begin{figure*} 
\gridline{\fig{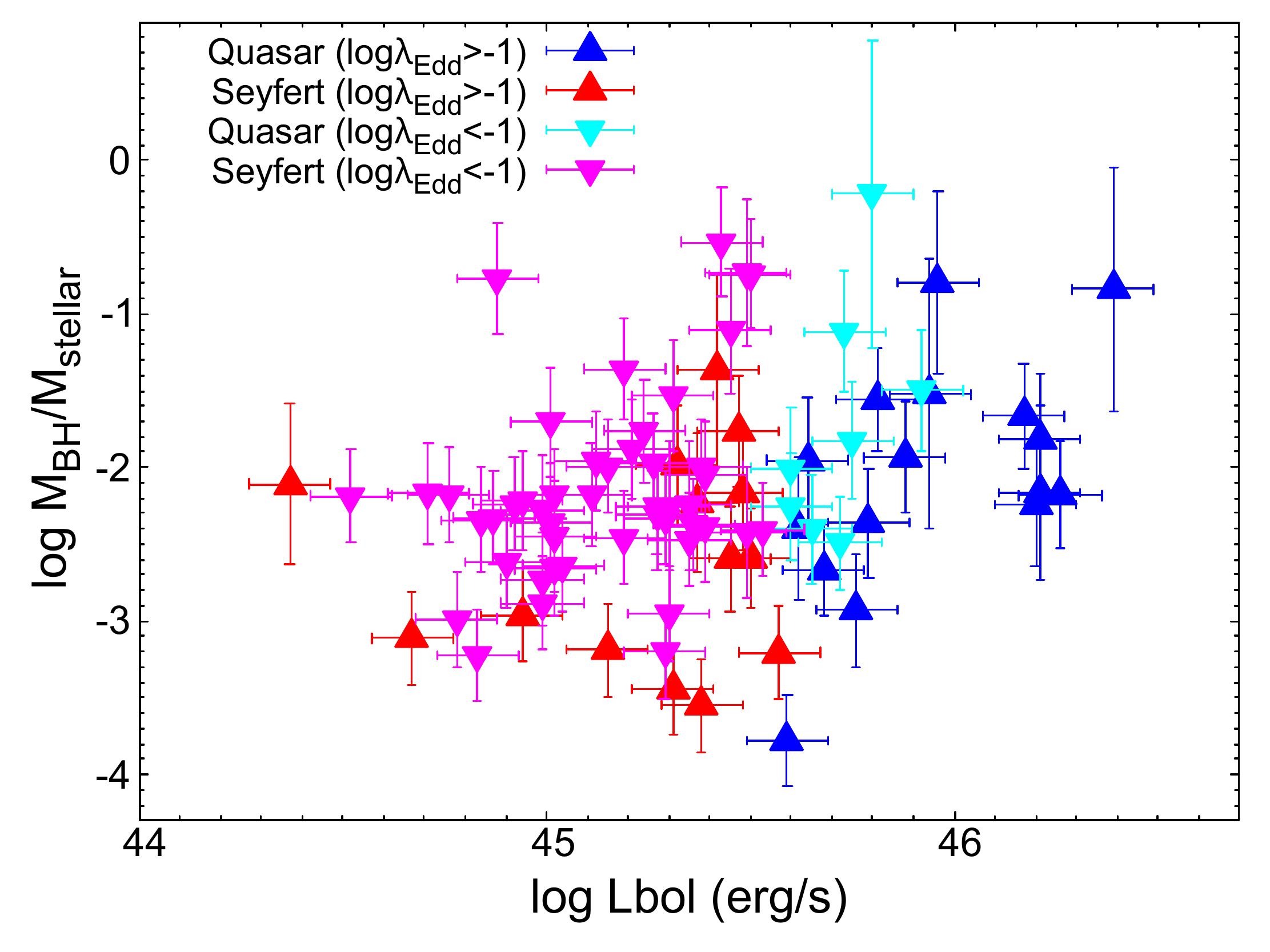}{0.45\textwidth}{(a)}
     \fig{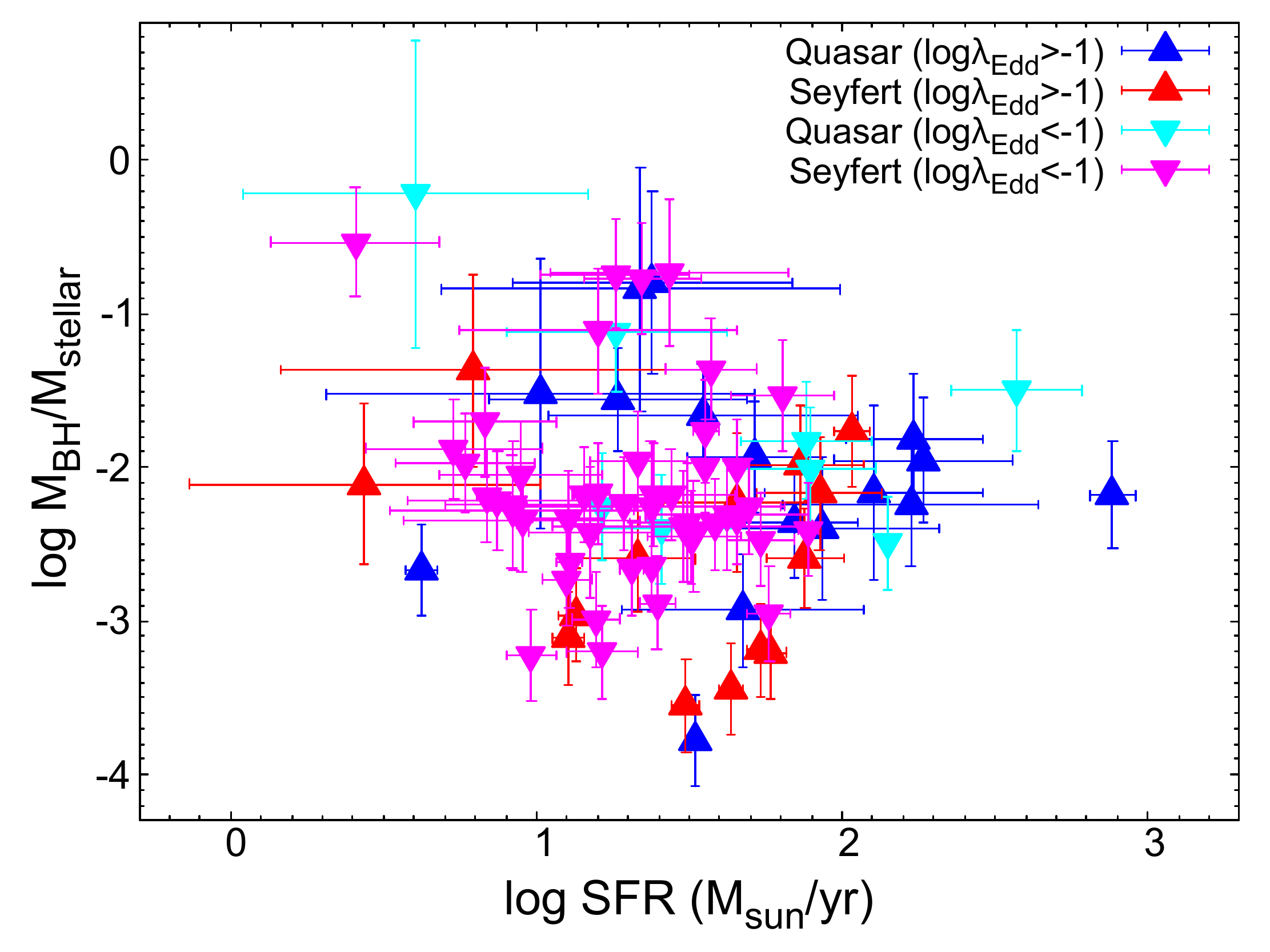}{0.45\textwidth}{(b)}
     }
\gridline{
\fig{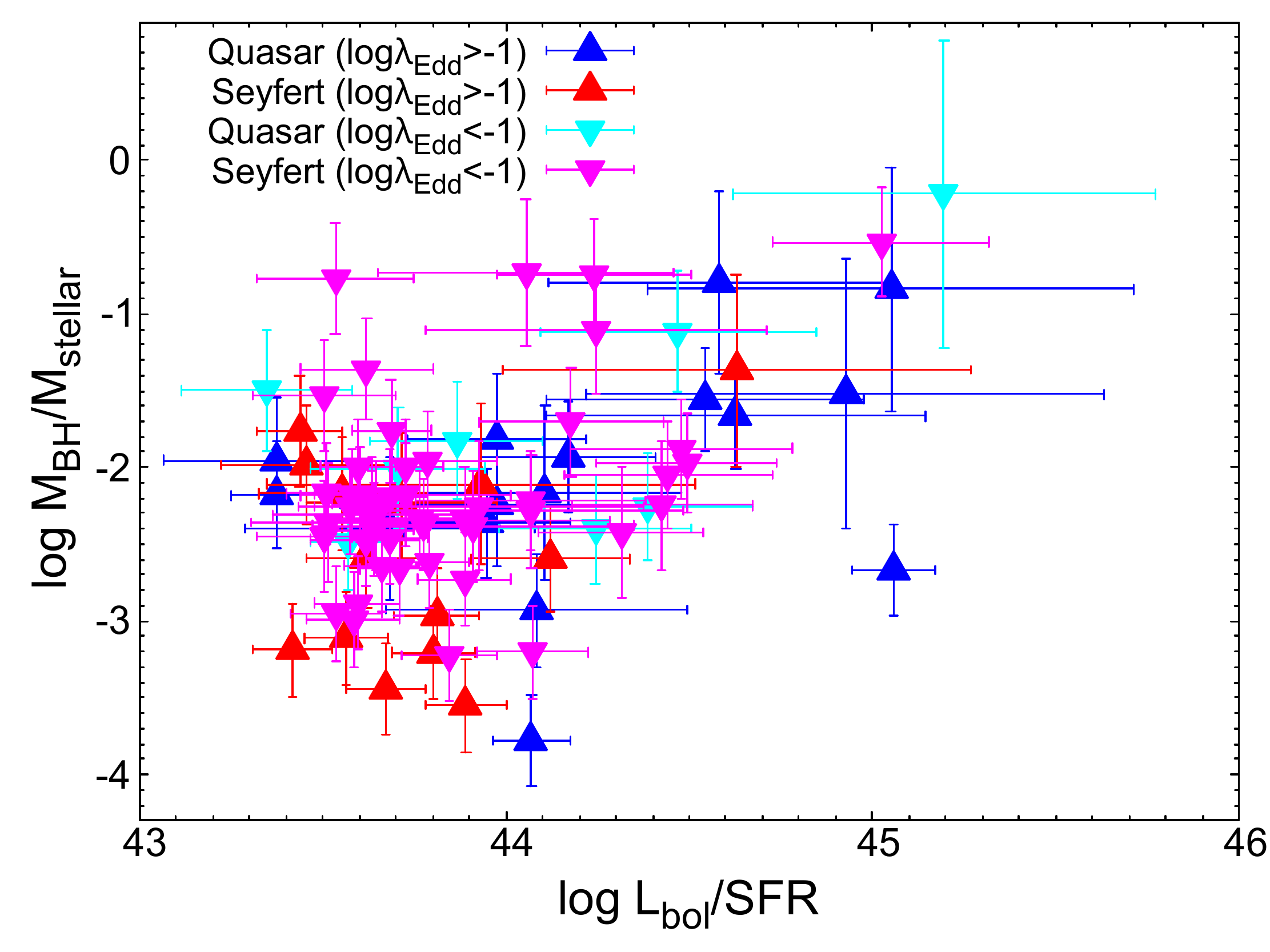}{0.45\textwidth}{(c)}
}
\caption{
$M_{\rm BH}/M_{\rm stellar}$ plotted against (a) $L_{\rm{bol}}$, (b)
SFR, and (c) $L_{\rm{bol}}$/SFR.
The symbols are the same as in Figure 2.
}
\label{fig:Lbol_sfr_vs_MBH_Mstellar}
\end{figure*}

The SMBH-to-stellar mass ratio gives us a hint on the evolutionary
scenario of our sample, in particular, whether the SMBHs grew earlier
or later than the galaxies. Figure~\ref{fig:Lbol_sfr_vs_MBH_Mstellar}
(a) and (b) plots the mass ratio against $L_{\rm bol}$ and SFR,
respectively. We obtain a correlation coefficient between $L_{\rm bol}$
and $M_{\rm{BH}}/M_{\rm{stellar}}$ of $r=0.324\pm0.138$ and that between
SFR and $M_{\rm{BH}}/M_{\rm{stellar}}$ of $r=-0.019\pm0.161$. 
The positive correlation of the SMBH-to-stellar mass ratio with the AGN luminosity but not with SFR
prefers an evolutionary scenario where star
formation precedes SMBH growth (as proposed by, e.g., \citealt{Ueda18})
in the moderately luminous AGN phase;
the reason why the correlation between SFR and
$M_{\rm{BH}}/M_{\rm{stellar}}$ is unseen may be explained by the
presence of the luminous quasars that show small SFR-to-$L_{\rm bol}$
ratios in the final stage of black hole mass growth. 
Our scenario predicts that the ratio of the black hole mass accretion rate to SFR increases with time during the moderately luminous AGN phase and hence should correlate with the SMBH-to-steller mass ratio. Figure~\ref{fig:Lbol_sfr_vs_MBH_Mstellar} (c) plots $M_{\rm{BH}}/M_{\rm{stellar}}$ against $L_{\rm bol}$/SFR. They show a weak positive correlation ($r=0.294\pm0.164$), which is in line with this picture.

\section{Conclusion}\label{sec:conclusion}

We have applied the {\tt X-CIGALE} code \citep{Yang20} to the far IR to far
UV SED of moderately luminous ($\log L_{\rm bol} \sim 44.5-46.5$), X-ray
selected broad-line AGNs at $z=1.18-1.68$ in the SXDF. The main
conclusions are summarized as follows.

\begin{itemize}

\item The mean ratio of the black hole mass to the total stellar mass
   for a Chabrier IMF (including that in the bulge and disk) is found
   to be --2.2, which is similar to the local SMBH-to-bulge mass
   ratio. This suggests that if the host galaxies of these
moderately luminous AGNs at $z\sim1.4$ are dominated by bulges, they
already established the local SMBH mass-bulge mass relation; if they
are dominated by disks, their SMBHs are overmassive relative to the
bulges.
However, a selection bias for luminous AGNs must be taken
   into account to discuss the properties of the whole galaxy
   population.
    
\item We find a good correlation between AGN bolometric luminosities and SFR
with ratios higher than that 
expected from the local $M_{\rm BH}-M_{\rm bulge}$ relation, 
suggesting that these AGNs are in a SMBH-growth dominant phase.
    
\end{itemize}

\acknowledgments

We acknowledge the anonymous referee for careful reading of this paper and constructive feedback.
We are deeply thankful to Prof. Veronique Buat, Dr. Guang Yang, and 
Prof. Denis Burgarella for helping to install {\tt X-CIGALE} code.
This work was financially supported by the Grant-in-Aid for Scientific
Research grant Nos. 20H01946 (Y.U.), 18J01050, and 19K14759 (Y.T.).

%

\vspace{5mm}







\end{document}